\documentstyle[12pt]{article}\textheight
235mm\textwidth 165mm \hoffset=-1.5cm
\newcommand{\bi}{\bibitem}
\newcommand{\be}{\begin{eqnarray}}
\newcommand{\ee}{\end{eqnarray}}
\newcommand{\nn}{\nonumber}
\catcode`\@=11
\def\lsim{\mathrel{\mathpalette\@versim<}}
\def\gsim{\mathrel{\mathpalette\@versim>}}
\def\@versim#1#2{\vcenter{\offinterlineskip
\ialign{$\m@th#1\hfil##\hfil$\crcr#2\crcr\sim\crcr } }}
\catcode`\@=12

\begin{document}
\pagestyle{empty}
\hspace*{10cm} \vspace{-3mm} MPI-PhT/96-71\\
\hspace*{10.7cm} \vspace{-3mm} KANAZAWA-96-17\\
\hspace*{10.7cm} \vspace{-3mm} IFUNAM-FT96-11\\
\hspace*{10.7cm} August 1996

\vspace{0.3cm}

\begin{center}
{\Large\bf  Perturbative Unification 
\vspace{-1mm}\\ of \vspace{-1mm}\\
Soft Supersymmetry--Breaking Terms}
\end{center} 

\vspace{1cm}

\begin{center}{\sc Jisuke Kubo}$\ ^{(1),*}$, 
{\sc Myriam
Mondrag{\' o}n}$\ ^{(2),*}$ \vspace{-1mm} and 
{\sc George Zoupanos}$\ ^{(3),***}$  
\end{center}
\begin{center}
{\em $\ ^{(1)}$ 
Physics Department, \vspace{-2mm} Faculty of Science, \\
Kanazawa \vspace{-2mm} University,
Kanazawa 920-11, Japan } \\
{\em $\ ^{(2)}$ Instituto de F{\' \i}sica,  \vspace{-2mm} UNAM,
Apdo. Postal 20-364,
M{\' e}xico 01000 D.F., M{\' e}xico}\\
{\em $\ ^{(3)}$ Max-Planck-Institut f\"ur Physik,
 Werner-Heisenberg-Institut \vspace{-2mm}\\
D-80805 Munich, Germany} \vspace{-3mm} \\ and  \vspace{-3mm} \\
{\em  Physics Department, \vspace{-2mm} National
Technical University, \\ GR-157 80 Zografou, 
Athens, Greece$\ ^{\dag}$}  
  \end{center}

\begin{center}
{\sc\large Abstract}
\end{center}

\noindent
Perturbative unification of soft supersymmetry--breaking
(SSB) parameters is proposed in 
 Gauge-Yukawa unified models.
The method,
which can be applied in any finite order in perturbation
theory,  consists in searching for renormalization
group invariant relations among the SSB 
parameters, which are
consistent with perturbative renormalizability. 
For the minimal Gauge-Yukawa unified model based on $SU(5)$ we find
that the low energy SSB sector
contains a single arbitrary parameter, the unified
gaugino mass.
Within a certain approximation we find 
 that the model predicts a
superpartner spectrum which is consistent with the experimental data.

\vspace*{1cm}
\footnoterule
\vspace*{2mm}
\noindent
$^{*}$Partially supported  by the Grants-in-Aid
for \vspace{-3mm} Scientific Research  from the Ministry of
Education, Science 
and Culture \vspace{-3mm}  (No. 40211213).\\
\noindent
$ ^{**}$
Partially supported \vspace{-3mm} by DGAPA under contract
IN110296. \\
\noindent
$ ^{***}$
Partially supported \vspace{-3mm} by C.E.C. project, 
CHRX-CT93-0319. \\
$ ^{\dag}$ Parmanent address

\newpage
\pagestyle{plain}
\section{Introduction}

The usual path chosen to reduce the independent parameters of a
theory is the introduction of a symmetry.  Grand Unified Theories
(GUTs)
 are
representative examples of such attempts.
A natural gradual 
extension of the GUT idea, which preserves their successes and
enhances the predictions, may be to attempt to relate 
the gauge and Yukawa
couplings, or in other words, to achieve Gauge-Yukawa
Unification (GYU).  

In recent papers,
we have proposed
 an alternative way to
achieve unification of couplings, 
which is based on the principles of reduction of couplings and
finiteness \footnote{Appropriate references may be
found in ref. \cite{kmoz1}.}.  These
principles, which are formulated in 
perturbation theory, are not explicit symmetry principles, although
they might imply symmetries.  
The former principle is based on the
existence of renormalization group
(RG) invariant relations among couplings, which 
do not necessarily result from a symmetry,
but nevertheless preserve
perturbative renormalizability.  Similarly, the latter 
one is based on
the fact that it is possible to find RG invariant
 relations among couplings
that keep finiteness in perturbation theory. 
We have found that various supersymmetric GYU models predict
mass values for the top and bottom quarks,
$M_t$ and $M_b$, which are consistent
with the experimental data, and that under certain
circumstances the different models can
be distinguished from each other if 
$M_t$ and $M_b$ can be more accurately 
measured \cite{kmoz2}.

The most arbitrary part of a phenomenologically
viable supersymmetric model is the breaking of supersymmetry.
It is widely believed that the breaking of supersymmetry
is soft whatever its origin is. 
If the model is coupled to supergravity, for instance,
one can compute in principle 
the soft supersymmetry--breaking (SSB) terms. In fact,
this is an attractive way to reduce the arbitrariness
of the SSB terms, where the gravitino mass $m_{2/3}$ defines
the scale of the supersymmetry--breaking \cite{nilles1}.

In this letter, we would like to extend
our  unification idea
to include the SSB sector. That is, we want to find
RG invariant relations among the SSB parameters
that are consistent with  perturbative 
renormalizability \footnote{A similar but different
idea has been recently proposed in refs. 
\cite{jack1,kazakov1}.}.
To be definite, we will consider  the minimal
SUSY $SU(5)$ model with the GYU in the third 
generation \cite{mondragon1}.
We will find that,
if one requires  the breaking of the
electroweak symmetry to occur in the desired manner,
 the SSB sector of the model
can be  completely fixed by the gaugino mass parameter $M$.
It will turn out that the asymptotic freedom in the 
SSB sector of the Gauge-Yukawa unified model can be achieved only
through the reduction of the SSB parameters.
We will then calculate
within a certain approximation the SSB parameters
of the minimal supersymmetric standard model (MSSM),
which will turn out to be consistent with the experimental data.
More details of our results will be published elsewhere.

\section{Formalism}

The reduction of couplings
 was originally formulated for massless theories
on the basis of the Callan-Symanzik equation \cite{zim1}.
The extension to theories with massive parameters
is not straightforward if one wants to keep
the generality and the rigor
on the same level as for the massless case;
one has 
to fulfill a set of requirements coming from
the renormalization group
equations,  the  Callan-Symanzik equations, etc.
along with the normalization
conditions imposed on irreducible Green's functions \cite{piguet1}.
There has been some progress in this direction \cite{zim2}. 
Here, to simplify the situation,  we would like  to
 assume  that
 a mass-independent renormalization scheme has been
employed so that all the  RG functions have only  trivial
dependencies of dimensional parameters. 

To be general, we consider  a renormalizable theory
which contain a set of $(N+1)$ dimension-zero couplings,
$\{\hat{g}_0,\hat{g}_1,\dots,\hat{g}_N\}$, a set of $L$ 
parameters with  dimension 
one, $\{\hat{h}_1,\dots,\hat{h}_L\}$,
and a set of $M$ parameters with dimension two,
$\{\hat{m}_{1}^{2},\dots,\hat{m}_{M}^{2}\}$. 
The renormalized irreducible vertex function 
satisfies the RG equation
\be
0 &=& {\cal D}\Gamma [~{\bf
\Phi}'s;\hat{g}_0,\hat{g}_1,\dots,\hat{g}_N;\hat{h}_1,\dots,\hat{h}_L;
\hat{m}^{2}_{1},\dots,\hat{m}^{2}_{M};\mu~]~,\\
{\cal D} &=& \mu\frac{\partial}{\partial \mu}+
~\sum_{i=0}^{N}\,\beta_i\,
\frac{\partial}{\partial \hat{g}_i}+
\sum_{a=1}^{L}\,\gamma_{a}^{h}\,
\frac{\partial}{\partial \hat{h}_a}+
\sum_{\alpha=1}^{M}\,\gamma^{m^2}_{\alpha}\frac{\partial}
{\partial \hat{m}_{\alpha}^{2}}+ ~\sum_{J}\,\Phi_I
\gamma^{\phi I}_{~~~J} \frac{\delta}{\delta \Phi_J}~.\nn
\ee
Since we assume a mass-independent renormalization scheme,
the $\gamma$'s have the form
\be
\gamma_{a}^{h} &=& \sum_{b=1}^{L}\,
\gamma_{a}^{h,b}(g_0,\dots,g_N)\hat{h}_b~,\nn\\
\gamma_{\alpha}^{m^2} &=&
\sum_{\beta=1}^{M}\,\gamma_{\alpha}^{m^2,\beta}(g_0,\dots,g_N)
\hat{m}_{\beta}^{2}+
\sum_{a,b=1}^{L}\,\gamma_{\alpha}^{m^2,a b}
(g_0,\dots,g_N)\hat{h}_a \hat{h}_b~,
\ee
where $\gamma_{a}^{h,b}, 
\gamma_{\alpha}^{m^2,\beta}$ and 
  $\gamma_{a}^{m^2,a b}$
are power series of the dimension-zero
couplings $g$'s in perturbation theory.

As in the massless case, we then look for 
 conditions under which the reduction of
parameters,
\be
\hat{g}_i &=&\hat{g}_i(g)~,~(i=1,\dots,N)~,\\
~\hat{h}_a &= &\sum_{b=1}^{P}\,
f_{a}^{b}(g) h_b~,~(a=P+1,\dots,L)~,\\
~\hat{m}_{\alpha}^{2} &= &\sum_{\beta=1}^{Q}\,
e_{\alpha}^{\beta}(g) m_{\beta}^{2}+
\sum_{a,b=1}^{P}\,k_{\alpha}^{a b}(g)
h_a h_b~,~(\alpha=Q+1,\dots,M)~,
\ee
is consistent with the RG equation (1),
where we assume that $g\equiv g_0$, $h_a \equiv
\hat{h}_a~~(1 \leq a \leq P)$ and 
$m_{\alpha}^{2} \equiv
\hat{m}_{\alpha}^{2}~~(1 \leq \alpha \leq Q)$ 
are independent parameters of the reduced theory.
We find  that the following set of
equations has to be satisfied:
\be
\beta_g\,\frac{\partial
\hat{g}_{i}}{\partial g} & =& \beta_i ~,~(i=1,\dots,N)~,\\
\beta_g\,\frac{\partial
\hat{h}_{a}}{\partial g}+\sum_{b=1}^{P}\gamma_{b}^{h}
\frac{\partial
\hat{h}_{a}}{\partial
h_b} &=&\gamma_{a}^{h}~,~(a=P+1,\dots,L)~,\\
\beta_g\,\frac{\partial
\hat{m}^{2}_{\alpha}}{\partial g}
+\sum_{a=1}^{P}\gamma_{a}^{h}
\frac{\partial
\hat{m}^{2}_{\alpha}}{\partial
h_a}+
\sum_{\beta=1}^{Q}\gamma_{\beta}^{m^2}
\frac{\partial
\hat{m}^{2}_{\alpha}}{\partial
m^{2}_{\beta}}
 &=&\gamma_{\alpha}^{m^2}~,~(\alpha=Q+1,\dots,M)~.
\ee
Using eq. (2) for $\gamma$'s, one finds that  eqs. (6)--(8)
reduce to 
\be
& &\beta_g\,\frac{d f_{a}^{b}}{d g}+
\sum_{c=1}^{P}\, f_{a}^{c} 
[\,\gamma_{c}^{h,b}+\sum_{d=P+1}^{L}\,
\gamma_{c}^{h,d}f_{d}^{ b}\,]
-\gamma_{a}^{h,b}-\sum_{d=P+1}^{L}\,
\gamma_{a}^{h,d}f_{d}^{ b}~=0~,\\
& &~(a=P+1,\dots,L; b=1,\dots,P)~,\nn\\
& &\beta_g\,\frac{d e_{\alpha}^{\beta}}{d g}+
\sum_{\gamma=1}^{Q}\, e_{\alpha}^{\gamma} 
[\,\gamma_{\gamma}^{m^2,\beta}+\sum_{\delta=Q+1}^{M}\,
\gamma_{\gamma}^{m^2,\delta}e_{\delta}^{\beta}\,]
-\gamma_{\alpha}^{m^2,\beta}-\sum_{\delta=Q+1}^{M}\,
\gamma_{\alpha}^{m^2,\delta}e_{\delta}^{\beta}~=0~,\\
& &~(\alpha=Q+1,\dots,M; \beta=1,\dots,Q)~,\nn\\
& &\beta_g\,\frac{d k_{\alpha}^{a b}}{d g}+2\sum_{c=1}^{P}\,
(\,\gamma_{c}^{h,a}+\sum_{d=P+1}^{L}\,
\gamma_{c}^{h,d}f_{d}^{a}\,)k_{\alpha}^{c b}+
\sum_{\beta=1}^{Q}\, e_{\alpha}^{\beta}
[\,\gamma_{\beta}^{m^2,a b}+\sum_{c,d=P+1}^{L}\,
\gamma_{\beta}^{m^2,c d}f_{c}^{a} f_{d}^{b}\nn\\
& &+2\sum_{c=P+1}^{L}\,\gamma_{\beta}^{m^2,c b}f_{c}^{a}+
\sum_{\delta=Q+1}^{M}\,\gamma_{\beta}^{m^2,\delta}
k_{\delta}^{a b} \,]-
[\,\gamma_{\alpha}^{m^2,a b}+\sum_{c,d=P+1}^{L}\,
\gamma_{\alpha}^{m^2,c d}f_{c}^{a} f_{d}^{b}\nn\\
& &+
2\sum_{c=P+1}^{L}\,\gamma_{\alpha}^{m^2,c b}f_{c}^{a}
+\sum_{\delta=Q+1}^{M}\,\gamma_{\alpha}^{m^2,\delta}
k_{\delta}^{a b} \,]~=0~,\\
& &(\alpha=Q+1,\dots,M; a,b=1,\dots,P)~.\nn
\ee
If these equations are satisfied, 
the irreducible vertex function of the reduced theory
\be
& &\Gamma_R [~{\bf
\Phi}'s; g; h_1,\dots,h_P; m^{2}_{1},
\dots,\hat{m}^{2}_{Q};\mu~]~\nn\\
&\equiv& \Gamma [~{\bf
\Phi}'s; g,\hat{g}_1(g),\dots,\hat{g}_N (g);
 h_1,\dots,h_P, \hat{h}_{P+1}(g,h),\dots,\hat{h}_L(g,h);\nn\\
& & m^{2}_{1},\dots,\hat{m}^{2}_{Q},\hat{m}^{2}_{Q+1}(g,h,m^2),
\dots,\hat{m}^{2}_{M}(g,h,m^2);\mu~] \ee 
has
the same renormalization group flow as the original one.

The requirement for the reduced theory to be perturbative
renormalizable means that the functions $\hat{g}_i $,
$f_{a}^{b} $, $e_{\alpha}^{\beta}$ 
and $k_{\alpha}^{a b}$, defined in eq. (3)--(5),  should have a power
series expansion in the primary coupling $g$:
\be
\hat{g}_{i} &=& g\,\sum_{n=0}^{\infty}
 \rho_{i}^{(n)} g^{n}~,~
f_{a}^{b}= g\sum_{n=0}^{\infty} \eta_{a}^{b~(n)} g^{n}~,\nn\\~
e_{\alpha}^{\beta} &= &\sum_{n=0}^{\infty} 
\xi_{\alpha}^{\beta~(n)} g^{n}~,~
k_{\alpha}^{a b }= \sum_{n=0}^{\infty} 
\chi_{\alpha}^{a b~(n)} g^{n}~,
\ee
To obtain the expansion coefficients, 
 we insert the power series ansatz
above into eqs. (6), (9)--(11)
and require that the equations are satisfied
 at each order in $g$. Note that the existence of a unique
power series solution is a non-trivial matter: It depends
on the theory as well as on the choice of the set of
independent parameters. In a concrete model we will consider
below, we will discuss this issue more in detail.

\section{Application to the minimal
SUSY $SU(5)$ GUT}

{\bf 3.1} {\em The model and its RG functions} \newline
The three
generations of quarks and leptons   are accommodated by 
three chiral superfields in
$\Psi^{I}({\bf 10})$ and $\Phi^{I}(\overline{\bf 5})$,
where $I$ runs over the three generations.
A $\Sigma({\bf 24})$ is used to break $SU(5)$ down to $SU(3)_{\rm C}
\times SU(2)_{\rm L} \times U(1)_{\rm Y}$,  and
$H({\bf 5})$ and $\overline{H}({\overline{\bf 5}})$
 to describe the
two Higgs superfields appropriate for 
electroweak symmetry breaking  \cite{sakai}.
The superpotential of the model is  \cite{sakai} \footnote{We
suppress the hat on the couplings from now on,
which was used in the previous section to distinguish
the independent parameters from the dependent ones.}
\be
W &=& \frac{g_{t}}{4}\,
\epsilon^{\alpha\beta\gamma\delta\tau}\,
\Psi^{(3)}_{\alpha\beta}\Psi^{(3)}_{\gamma\delta}H_{\tau}+
\sqrt{2}g_b\,\Phi^{(3) \alpha}
\Psi^{(3)}_{\alpha\beta}\overline{H}^{\beta}+
\frac{g_{\lambda}}{3}\,\Sigma_{\alpha}^{\beta}
\Sigma_{\beta}^{\gamma}\Sigma_{\gamma}^{\alpha}+
g_{f}\,\overline{H}^{\alpha}\Sigma_{\alpha}^{\beta} H_{\beta}\nn\\
& &+ \frac{\mu_{\Sigma}}{2}\,
\Sigma_{\alpha}^{\gamma}\Sigma_{\gamma}^{\alpha}+ 
+\mu_{H}\,\overline{H}^{\alpha} H_{\alpha}~, 
\ee
where $\alpha,\beta,\ldots$ are the $SU(5)$
indices, and we have suppressed the Yukawa couplings of the
first two generations.
The Lagrangian containing the SSB terms
is
\be
-{\cal L}_{\rm soft} &=&
m_{H_u}^{2}{\hat H}^{* \alpha}{\hat H}_{\alpha}
+m_{H_d}^{2}
\hat{\overline {H}}^{*}_{\alpha}\hat{\overline {H}}^{\alpha}
+m_{\Sigma}^{2}{\hat \Sigma}^{\dag~\alpha}_{\beta}
{\hat \Sigma}_{\alpha}^{\beta}
+\sum_{I=1,2,3}\,[\,
m_{\Phi^I}^{2}{\hat \Phi}^{* ~(I)}_{\alpha}{\hat \Phi}^{(I)\alpha}
\nn\\
& &+\,m_{\Psi^I}^{2}{\hat \Psi}^{\dag~(I)\alpha\beta}
{\hat \Psi}^{(I)}_{\beta\alpha}\,]
+\{ \,
 \frac{1}{2}M\lambda \lambda+
B_H\hat{\overline {H}}^{\alpha}{\hat H}_{\alpha}
+B_{\Sigma}{\hat \Sigma}^{\alpha}_{\beta}
{\hat \Sigma}_{\alpha}^{\beta}
+h_{f}\,\hat{\overline{H}}^{\alpha}
{\hat \Sigma}_{\alpha}^{\beta} {\hat H}_{\beta}\nn\\
& &+\frac{h_{\lambda}}{3}\,{\hat \Sigma}_{\alpha}^{\beta}
{\hat \Sigma}_{\beta}^{\gamma}{\hat \Sigma}_{\gamma}^{\alpha}+
\frac{h_{t}}{4}\,
\epsilon^{\alpha\beta\gamma\delta\tau}\,
{\hat \Psi}^{(3)}_{\alpha\beta}
{\hat \Psi}^{(3)}_{\gamma\delta}{\hat H}_{\tau}+
\sqrt{2}h_{b}\,{\hat \Phi}^{(3) \alpha}
{\hat \Psi^{(3)}}_{\alpha\beta}\hat{\overline{H}}^{\beta}
+\mbox{h.c.}\, \}~,
\ee
where a hat is used to denote the scalar
component of each chiral superfield.

The RG functions of this model may be found 
in refs. \cite{mondragon1,polonsky1,kazakov1}, and
we employ the usual
 normalization  of the RG functions, 
$d {\rm A}/d \ln \mu ~=~
[\beta^{(1)}(A) ~~\mbox{or}~~   \gamma^{(1)}(A)]/16
\pi^2+\dots$, where $\dots$ are higher orders,
and $\mu$ is the renormalization scale: 
\be
\beta^{(1)}(g) &=& -3 g^3~,~
\beta^{(1)}(g_t) = 
[\,-\frac{96}{5}\,g^2+9\,g_{t}^{2}+\frac{24}{5}\,g_{f}^{2}+
4\,g_{b}^{2}\,]\,g_{t}~,\nn\\
\beta^{(1)}(g_b) &=& 
[\,-\frac{84}{5}\,g^2+3\,g_{t}^{2}+\frac{24}{5}\,g_{f}^{2}+
10\,g_{b}^{2}\,]\,g_{b}~,\nn\\
\beta^{(1)}(g_\lambda) &=& 
[\,-30\,g^2+\frac{63}{5}\,g_{\lambda}^2+3\,g_{f}^{2}
\,]\,g_{\lambda}~,\nn\\
\beta^{(1)}(g_f) &=& 
[\,-\frac{98}{5}\,g^2+3\,g_{t}^{2}
+4\,g_{b}^{2}
+\frac{53}{5}\,g_{f}^{2}+\frac{21}{5}\,g_{\lambda}^{2}
\,]\,g_{f}~,~
\gamma^{(1)}(M) = -6g^2 \,M~,\nn\\
\gamma^{(1)}(\mu_{\Sigma}) &=& [\, -20g^2 +2g_{f}^{2}
+\frac{42}{5} g_{\lambda}^{2}\,]\,\mu_{\Sigma} ~,~
\gamma^{(1)}(\mu_H) =
[\, -\frac{48}{5}g^2 +\frac{48}{5}g_{f}^{2}
+4 g_{b}^{2}+3g_{t}^{2}\, ]\,\mu_H ~,\nn\\
\gamma^{(1)}(B_H) &=& 
[\, -\frac{48}{5}g^2 +\frac{48}{5}g_{f}^{2}
+4 g_{b}^{2}+3g_{t}^{2} \,]\, B_H \nn\\
& &+
[\,\frac{96}{5}g^2 M+\frac{96}{5}h_{f}g_{f}
+8 g_b h_{b}+6 g_t h_{t}]\, \mu_H ~, \nn\\
\gamma^{(1)}(B_{\Sigma}) &=& 
[\, -20 g^2 +2g_{f}^{2}
+\frac{42}{5} g_{\lambda}^{2} \,]\, B_{\Sigma}+
[\,40 g^2 M+4 h_{f}g_{f}
+\frac{84}{5} g_{\lambda} h_{\lambda}]\, \mu_{\Sigma} ~, \nn\\
\gamma^{(1)}(h_t) &=& 
[\,-\frac{96}{5}\,g^2+9\,g_{t}^{2}+\frac{24}{5}\,g_{f}^{2}+
4\,g_{b}^{2}\,]\,h_t\nn\\
& &+[\, \frac{192}{5} M g^2+
18 h_t g_t+8 h_b g_b +\frac{48}{5}h_f g_f\,] \, g_t~,\nn\\
\gamma^{(1)}(h_b) &=& 
[\,-\frac{84}{5}\,g^2+3\,g_{t}^{2}+\frac{24}{5}\,g_{f}^{2}+
10\,g_{b}^{2}\,]\, h_b\nn\\
& &+[\, \frac{168}{5} M g^2+
6 h_t g_t+20 h_b g_b +\frac{48}{5}h_f g_f\,]\, g_b~~,\\
\gamma^{(1)}(h_{\lambda}) &=& 
[\,-30\,g^2+\frac{63}{5}\,g_{\lambda}^2+3\,g_{f}^{2}
\,]\,h_{\lambda}
+[\, 60 M g^2+
\frac{126}{5}h_{\lambda} g_{\lambda}
 +6h_f g_f\,] \,g_{\lambda}~,\nn\\
\gamma^{(1)}(h_f) &=& 
[\,-\frac{98}{5}\,g^2+3\,g_{t}^{2}
+4\,g_{b}^{2}
+\frac{53}{5}\,g_{f}^{2}+\frac{21}{5}\,g_{\lambda}^{2}
\,]\, h_f\nn\\
& &~+[\, \frac{196}{5} M g^2+6 h_t g_t+8 h_b g_b+
\frac{42}{5}h_{\lambda} g_{\lambda}
 +\frac{106}{5}h_f g_f\,] \, g_{f} ~,\nn\\
\gamma^{(1)}(m_{H_d}^{2}) &=& 
 -\frac{96}{5} g^2 M^{2}+\frac{48}{5}g_{f}^{2}(m_{H_u}^{2}+
m_{H_d}^{2}+m_{\Sigma}^{2})
\nn\\
& &+
8 g_{b}^{2}(
m_{H_d}^{2}+m_{\Psi^3}^{2}+m_{\Phi^3}^{2})+
\frac{48}{5} h_{f}^{2}+8 h_{b}^{2}~, \nn\\
\gamma^{(1)}(m_{H_u}^{2}) &=& 
 -\frac{96}{5} g^2 M^{2}+\frac{48}{5}g_{f}^{2}(m_{H_u}^{2}+
m_{H_d}^{2}+m_{\Sigma^3}^{2})+
6 g_{t}^{2}(
m_{H_u}^{2}+2m_{\Psi^3}^{2})+
\frac{48}{5} h_{f}^{2}+6 h_{t}^{2}~, \nn\\
\gamma^{(1)}(m_{\Sigma}^{2}) &=& 
 -40 g^2 M^{2}+2g_{f}^{2}(m_{H_u}^{2}+
m_{H_d}^{2}+m_{\Sigma}^{2})
+
\frac{126}{5} g_{\lambda}^{2}m_{\Sigma}^{2}+
2 h_{f}^{2}+\frac{42}{5} h_{\lambda}^{2}~,\nn\\
\gamma^{(1)}(m_{\Phi^3}^{2}) &=& 
 -\frac{96}{5} g^2 M^{2}+
8 g_{b}^{2}(
m_{H_d}^{2}+m_{\Psi^3}^{2}+m_{\Phi^3}^{2})+
8 h_{b}^{2} ~,\nn\\
\gamma^{(1)}(m_{\Psi^3}^{2}) &=& 
 -\frac{144}{5} g^2 M^{2}+6g_{t}^{2}(
m_{H_u}^{2}+2m_{\Psi^3}^{2})+
4 g_{b}^{2}(
m_{H_d}^{2}+m_{\Psi^3}^{2}+m_{\Phi^3}^{2})+
6h_{t}^{2}+4  h_{b}^{2}~,\nn\\
\gamma^{(1)}(m_{\Phi^{1,2}}^{2}) &=& 
-\frac{96}{5} g^2 M^{2} ~,~
\gamma^{(1)}(m_{\Psi^{1,2}}^{2}) = 
-\frac{144}{5} g^2 M^{2} ~,\nn
\ee
where $g$ stands for the gauge coupling.

\vspace{0.3cm}

\noindent
{\bf 3.2} {\em The reduction solution} \newline
 We require that the reduced theory
should contain the minimal number of the SSB parameters that are
consistent with   perturbative renormalizability. We will find
that the set of the  perturbatively unified SSB parameters
significantly differ from the so-called universal SSB parameters.

Without loss of generality, 
one can assume that the gauge coupling $g$ is the primary coupling.
Note that the reduction solutions in the dimension-zero sector
is independent
of the dimensionfull sector (under the assumption of a
mass independent renormalization scheme).
It has been found \cite{mondragon1} that there exist
two asymptotically free (AF) solutions that make 
a Gauge-Yukawa Unification possible in the present model:
\be
a & :& g_t=\sqrt{\frac{2533}{2605}} g +0(g^3)~,~
g_b=\sqrt{\frac{1491}{2605}} g +0(g^3)~,~
g_{\lambda}=0~,~
g_f=\sqrt{\frac{560}{521}} g +0(g^3)~,\nn\\
b & :& g_t=\sqrt{\frac{89}{65}} g +0(g^3)~,~
g_b=\sqrt{\frac{63}{65}} g +0(g^3)~,~
g_{\lambda}=0~,~g_f=0~,
\ee
where the higher order terms denote uniquely computable
power series in $g$.
It has been also found that the two solutions in (17)
describe the boundaries of an
asymptotically free RG-invariant surface in the space
of the couplings, on which $g_{\lambda}$ and $g_f$ 
can be different from zero. This observation 
has enabled us to obtain a partial reduction of couplings
for which the $g_{\lambda}$ and $g_f$ can be treated
as (non-vanishing) independent parameters without loosing 
AF.
Later we have found \cite{kmoz2} that the region on the
AF surface consistent with the 
proton decay constraint
has to be very close to the solution $a$.
Therefore, we assume in the following discussion
that we are exactly at the boundary defined by the solution $a$
\footnote{How to  go away slightly from this boundary 
will be discussed
elsewhere. Note that
$ g_{\lambda}=0 $ is inconsistent, but $g_{\lambda} < \sim 0.005$
has to be fulfilled to satisfy the proton 
decay constraint \cite{kmoz2}.
We expect that the inclusion of a small $g_{\lambda} $ will
not affect the prediction of the perturbative unification of
the SSB parameters.}.

In the dimensionful sector, we seek the reduction
of the parameters in the form (4) and (5).
First, one can realize that the
supersymmetric mass parameters, $\mu_{\Sigma}$ and 
$\mu_H$, and the gaugino mass parameter
$M$ cannot be reduced; that is, there
is no solution in the desired form. Therefore,  they
should be treated as independent parameters.
We find the following lowest order
reduction solution:
\be
B_H &=& \frac{1029}{521}\,\mu_H M~,~
B_{\Sigma}=-\frac{3100}{521}\,\mu_{\Sigma} M~,
\ee
\be
h_t &=&-g_t\,M~,~h_b =-g_b\,M~,
~h_f =-g_f\,M~,~h_{\lambda}=0~,\nn\\
m_{H_u}^{2} &=&-\frac{569}{521} M^{2}~,~
m_{H_d}^{2} =-\frac{460}{521} M^{2}~,
~m_{\Sigma}^{2} = \frac{1550}{521} M^{2}~,\nn\\
m_{\Phi^3}^{2} & = &\frac{436}{521} M^{2}~,~
m_{\Phi^{1,2}}^{2} =\frac{8}{5} M^{2}~,~
m_{\Psi^3}^{2} =\frac{545}{521} M^{2}~,~
m_{\Psi^{1,2}}^{2} =\frac{12}{5} M^{2}~.
\ee
So, the gaugino mass parameter $M$ plays 
a similar role as the gravitino mass $m_{2/3}$ in
supergravity coupled GUTs and characterizes
the scale of the supersymmetry--breaking.

In addition to the $\mu_{\Sigma}$,
$\mu_H$ and $M$, it is possible to include  also 
$B_H$ and $B_{\Sigma}$ as independent parameters
without changing the one-loop 
reduction solution (19).

\vspace{0.3cm}

\noindent
{\bf 3.3} {\em Uniqueness of the reduction}\newline
We next address the question of whether the 
lowest-order solution given in (18) and (19)
can
be uniquely extended to a power series solution in higher orders.
In  ref. \cite{mondragon1}, the uniqueness in the dimension-zero
sector is proved, and so we assume here that  the reduction 
in this sector has been performed.

Let us begin with  the case of  $h_a~~ (a=t,b,f)$.
We prove the uniqueness by induction;
we assume that the reduction is unique to $O(g^{n-1})$ and
show that the expansion coefficients
in the next order can be uniquely calculated.
We then insert the ansatz
\be
h_a &=&-g_a M+\dots+g g^n   \eta_{a}^{(n)} M~,~
a=t,b,f~,
\ee
along with the  solution $a$
in the dimension-zero sector (17), 
into the reduction equation (9) using eq. (13).
Then collecting
terms of $O(g^{n+3})$, one obtains 
$\sum_{c=t,b,f}\,L_{a c}(n) \eta_{c}^{(n)}
 = \cdots$,
where $\cdots$ in the r.h. side is known by assumption.
One  finds that  the determinant,
\be
\det L(n) &=&
\frac{38423832921}{6786025} + 
\frac{21646499373}{6786025}n + \frac{1423971}{2605}n^2 + 
  27 n^3~,
\ee
 for integer 
$n > 0$ never vanishes, implying that the expansion
coefficients $\eta_{a}^{(n)}$ can be uniquely calculated.
Since the one-loop reduction (19) is unique, the $\eta$'s
exist uniquely to any finite order.

The uniqueness in 
the dimension-two sector proceeds similarly.
Note that the uniqueness of the expansion coefficients for
$B_H$, $B_{\Sigma}$, $m_{\Phi^{1,2}}^{2}$ and
$m_{\Psi^{1,2}}^{2}$ can be easily shown, because
their one-loop anomalous dimensions are such that
there exists no mixing among the coefficients
(see eq. (16)). In the case of $m_{\alpha}^{2}~
(\alpha=H_d,H_u,\Sigma,\Phi^3,\Psi^3)$, we have to do a
similar investigation as for the $h$'s.
So we start with
$
m_{\alpha}^{2} = \xi_{\alpha}^{(0)} M^{2}+\dots
+g^n
\xi_{\alpha}^{(n)} M^{2}$,
where \footnote{As for the case of $h_a$'s, we have assumed that
the $\gamma (m^{2})$'s are independent of the supersymmetric
mass parameters $\mu_H$ and $\mu_{\Sigma}$.}
the lowest order coefficients 
$\xi_{\alpha}^{(0)}$ can be read off from (19),
and we assume that the lower order 
terms denoted by $\dots$ are
known. After some algebraic calculations, one finds
that 
the $\xi_{i}^{(n)}$ also can be uniquely calculated 
to any finite 
order \footnote{The approach of unifying the SSB parameters 
of ref. \cite{jack1} is based on a condition
on the anomalous dimensions (the $P=Q/3$ condition).
This condition is more restrictive than simply requiring
the complete reduction of parameters, because
the number of the anomalous dimensions usually exceeds
that of parameters. It has turned out to be very difficult to
satisfy the  $P=Q/3$ condition in higher orders in 
non-finite theories \cite{jack2}.}.

\vspace{0.3cm}

\noindent
{\bf 3.4} {\em Asymptotic freedom (AF) and the stability of
the reduction solution} \newline
If a reduction solution
is unstable, the aysmptotic freedom requirement
and the requirement on 
 a power series reduction solution are equivalent in general.
In what follows, we show that the reduction solution
(19) is an unstable asymptotically free solution and
exhibits the 
Pendleton--Ross infrared fixed point \cite{pendleton1}.
That is, the AF requirement forces
all the $h_a$'s and $m_{\alpha}^{2}$'s to be reduced according to 
the reduction solution (19).
On contrary,  $B_H$ and $B_{\Sigma}$
behave asymptotically free, and
their reduction solution (18) will turn out to be  stable.
To see these, we first  derive the asymptotic behavior of
the independent parameters, $\mu_{\Sigma}$, $\mu_{H}$
and $M$:
\be
\mu_{\Sigma} & \sim & g^{3100/1653}~,~
\mu_{H}  \sim  g^{-1029/521}~,~
M  \sim  g^{2}~~\mbox{as}~~g \to 0~,
\ee
where we have used eq. (17) and
$ d/d \ln \mu = (-3 g^3 +O(g^5))d/d g$.
So, the $  \mu_{H} $ does not vanish asymptotically.
Note, however, that thanks to the AF in the Gauge-Yukawa
sector the asymptotic behavior given in (22)  becomes
exact in the ultraviolet limit.
Moreover, in a mass independent renormalization scheme
(which we are assuming throughout),
the supersymmetric mass parameters 
$\mu_H$ and $\mu_{\Sigma}$ do not enter
in the anomalous dimensions for $h$'s 
and $m^{2}$'s \cite{gates1} so that
the investigation below is not affected by the bad 
asymptotic behavior of $\mu_{H} $.
To proceed, we introduce 
$\tilde{h}_a \equiv h_a/M$ and $
\tilde{m}_{\alpha}^{2} \equiv 
m_{\alpha}^{2}/M^{2}$,
and consider a solution near the reduction solution (19):
$\tilde{h}_a (g) =-g_a +\Delta^{h}_{a}(g)~,~a=t,b,f$.
Then we derive from eq. (7) 
the linearized equations 
\be
\frac{d \Delta^{h}_{a}(g)}{d g}&=&
 \sum_{c=t,b,f} Y_{a c}\Delta^{h}_{c}(g)/g~.
\ee
The asymptotic behavior
of the system is dictated by the eigenvalues
of the matrix $Y$, and one finds that
the three basis vectors ${\bf  v}^{h}_{i}(g)$ behave like
\be
{\bf  v}^{h}_{i} &\sim & g^{\lambda_i}~,~
\lambda_i=-11.64\dots, -4.98\dots, -3.61\dots~,
\ee
as $g \to 0$, where the $\lambda_i$'s are the eigenvalues of $Y$,
implying  that the reduction solution for
$h_a$'s is ultraviolet unstable.
One also sees that AF requires the $h_a$'s to be reduced 
because
$M \sim g^2$ as $g \to 0$.

The $m^{2}$-sector can be discussed similarly.
Assuming that 
$\tilde{m}_{\alpha}^{2} (g) =
\xi_{(0)}^{5}  +\Delta^{m^2}_{\alpha}(g)~,~
\alpha=H_d,H_u,\Sigma,\Phi^{1,2,3},\Psi^{1,2,3}$,
and that the $h_a$'s are reduced, we find that
the eigenvalues of the matrix $Z$ which enters in
the linearized equations, 
$ d \Delta^{m^2}_{\alpha}(g)/d g =
\sum_{\beta=H_d,H_u,\Sigma,\Phi^3,\Psi^3}
Z_{\alpha\beta}\Delta^{m^2}_{\beta}(g)/g$,
are given by
$(\,-14.64\dots, -7.98\dots,  -6.61\dots,-4,-4,-4,-4\,)$.
Therefore,  the reduction solution for
$m_{\alpha}^{2}$'s is also ultraviolet unstable, and one,
moreover, sees that the AF of
$m_{\alpha}^{2}$'s is ensured  only by the reduction (19)
because $M^{2} \sim g^4~~\mbox{as}~~g \to 0$.

As for $B_H$ and $B_{\Sigma}$, we find that
as $g \to 0$,
\be
B_H &\simeq &\frac{1029}{521}\mu_H M+c_H\,g^{1.97\dots}~,~
B_{\Sigma} \simeq 
-\frac{3100}{521}\mu_{\Sigma} M+c_{\Sigma}
\,g^{0.64\dots}~
\ee
near the reduction solution,
where $c$'s are integration constants.
Therefore, the $B$'s are asymptotically free
($\mu_H M \sim g^{0.024\dots}~,
~\mu_{\Sigma} M \sim g^{3.8\dots}$),
 and so the reduction solution for the $B$'s
are  asymptotically stable. This is  good news, because,
as we will see later, the  reduction solution
(19) including (18) is not consistent 
with the radiative breaking
of the electroweak symmetry at  low energy.
To make the radiative breaking possible, we have to 
treat $B_H$ as an independent parameter.
But, as we have just seen, this can be done without
loosing AF of the model.

The solution (19) exhibits the 
one-loop infrared fixed point, which therefore could be used for
the infrared-fixed-point approach \cite{lanross}.
This approach 
is based on the assumption
that infrared fixed points found in first order 
in perturbation theory 
persist in higher orders
and that the ratio of the compactification scale $\Lambda_{\rm C}$
(or the Planck scale $M_{\rm P}$) to
$M_{\rm GUT}$ is large enough for 
various parameters 
to come very close to their infrared values when running
from $\Lambda_{\rm C}$  down to $M_{\rm GUT}$.
Therefore, this approach may yield similar results to ours,
because the reduction solution in  one-loop order (19)
is the infrared fixed point.
Here we would like to see
how fast the desired infrared fixed point can be
approached in our concrete  model.

To this end, we assume that $h_a$, $a=t,b,f$ and
$m_{\alpha}^{2}$, $\alpha=
H_d,H_u,\Sigma,\Phi^{1,2,3}, \Psi^{1,2,3} $ vanish at 
$\Lambda_{\rm C}$, while we treat $M$ as independent.
The one-loop evolution of $ m_{\Phi^{1,2}}^{2} $ and 
$ m_{\Psi^{1,2}}^{2} $ can be discussed
analytically: 
\be
\frac{m_{\Phi^{1,2}}^{2}}{M^{2}} &=&
\frac{8}{5}+c_{\Phi^{1,2}} g^{-4}~,~
\frac{m_{\Psi^{1,2}}^{2}}{M^{2}} =
\frac{12}{5}+c_{\Psi^{1,2}} g^{-4}~,
\ee
where $c$'s are integration constants.
Imposing the above mentioned boundary condition at
$\Lambda_{\rm C}$, one finds at $M_{\rm GUT}$
\be
\frac{m_{\Phi^{1,2}}^{2}}{M^{2}} &\simeq&
0.25, 0.35, 0.52   ~,~\frac{m_{\Psi^{1,2}}^{2}}{M^{2}}\simeq
0.37, 0.53, 0.79~~\mbox{for}~~\frac{\Lambda_{\rm C}}{M_{\rm GUT}}
=   10^2, 10^3, 10^5  ~,
\ee
respectively, where we have used $\alpha=g^2/4\pi=0.04$
at $M_{\rm GUT}$.  Unfortunately, we see that
 the infrared fixed point,
$1.6$ and $2.4$, is quite far from the
approached points. We have checked numerically that this 
also holds for the other SSB parameters.

\vspace{0.3cm}

\noindent
{\bf 3.5} {\em Prediction}\newline
Since the $SU(5)$ symmetry is spontaneously broken
at $M_{\rm GUT}$, the reduction relations (17)-(19)
exhibit a boundary condition on the
gauge and Yukawa couplings and also on the SSB parameters
at this energy scale \footnote{Here we examine the evolution of these
parameters according to their renormalization group equations
 in two-loop order for the  
gauge and Yukawa couplings and 
in one-loop order for the SSB parameters.}.
To make our unification idea and its consequence transparent,
we shall make an oversimplifying assumption
that below $M_{\rm GUT}$ their
evolution is  governed by the MSSM and that 
there exists a unique threshold
$M_{\rm SUSY}$,
which we identify with $M$, for all superpartners of the
MSSM, so that below $M_{\rm SUSY}$ the standard model (SM) is the
correct effective theory. 
We  recall that
it is most convenient to fix
$\tan\beta$ through the matching condition on the Yukawa couplings
at $M_{\rm SUSY}$ in the Gauge-Yukawa Unification
 scenario \cite{mondragon1,kmoz2}. That is,
the Higgs sector is partly fixed by the dimension-zero sector.
This is the reason why
the complete reduction in the dimensionfull sector,
defined by (18) and (19), is inconsistent with the radiative 
breaking of the electroweak symmetry, as we will see below.

Since we are not 
stressing the accuracy of the approximation, we assume that
 the potential of the MSSM 
at $\mu=M$ takes the tree-level form.
The  minimization of the potential yields
 two conditions at $M_{\rm SUSY}$ \cite{ibanez1},
\be
0 &=& m_{H_d}^{2}-m_{H_u}^{2}+ M_{Z}^{2}
\frac{1-\tan^2\beta}{1+\tan^2\beta}+
B_H\frac{\tan^2\beta-1}{\tan\beta}~,\\
0 &=&
2\mu_{H}^{2}+m_{H_d}^{2}+m_{H_u}^{2}+
B_H\frac{\tan^2\beta+1}{\tan\beta}~,
\ee
where
$\tan\beta=v_2/v_1  ~,~M_Z=
(1/2)\sqrt{(3 g_{1}^{2}/5
+g_{2}^{2})( v_{1}^{2}+v_{2}^{2})}~,
~v_{1,2}=(1/\sqrt{2})<\hat{H}_{d,u}>$.
Using the
unification condition given by (18) and (19)
under the assumption that
$M_Z $ and $\tan\beta $ at $M_{\rm SUSY}$ are
given, these two conditions  
could fix the $M$ and $\mu_H$ at $M_{\rm GUT}$.
Unfortunately, this is not the case.
We have numerically checked that  the
unification condition given by (17)-(19)
does not satisfy eqs. (28) and (29).
Therefore, we have to treat one of $m_{H_u}$, 
$m_{H_d}$ and $B_H$ as an independent
parameter to make the radiative breaking at $M_{\rm SUSY}$
possible. From the discussion of sect. 3.4 it is clear that 
the most natural choice is $B_H$, because this is the unique possibility
to keep AF. In addition, the lowest order
unification condition (19) remains the same;
otherwise it would be modified.

We use
\be
\alpha_1(M_Z) &=&0.0169~,~
\alpha_2(M_Z) =0.0337~,~
\alpha_{\tau}(M_Z) =8.005\times 10^{-6}~
\ee
as input parameters and fix
$M_{\rm SUSY}=M$ at $500$ GeV. 
Then the prediction from the Gauge-Yukawa
Unification (17) is:
\be
M_{t} &\simeq& 1.8 \times 10^2 ~\mbox{GeV}~,~
M_{b} \simeq 5.4 ~\mbox{GeV}~,~\alpha_3 (M_Z)\simeq 0.12~,\nn\\
M_{\rm GUT} &\simeq& 1.7\times 10^{16}~\mbox{GeV}~,
\alpha_{\rm GUT} \simeq 0.040~,
~\tan\beta (M_{\rm SUSY})\simeq 48~,
\ee
where $M_{t}$ and $M_b$ are the physical top and bottom quark masses.
These values suffer from corrections coming from different sources
such as threshold effects, which are partly taken into account and
estimated in ref. \cite{kmoz2}.
In table 1, we show
 the prediction of the SSB parameters.  

\begin{center}
\begin{tabular}{|c||c|c||c|}
 \hline
$M_1$ (TeV) &
0.22 &
$m_{L_3}^{2}$ (TeV$ ^2$)  &
0.30 \\ \hline
$M_2$ (TeV) &
0.42 &
$m_{\tau}^{2}$ (TeV$ ^2$)  &
0.23 
\\ \hline
$M_3 $ (TeV) &
1.2 &
$m_{Q_3}^{2}$ (TeV$ ^2$)   &
1.1 
\\ \hline
$h_t $ (TeV) & 
-0.89 &
$m_{b}^{2}$ (TeV$ ^2$)   &
0.95
\\ \hline
$h_b $ (TeV) &
-0.88 &
$m_{t}^{2}$ (TeV$ ^2$) &
0.93 
\\ \hline
$h_{\tau} $ (TeV) &
-0.12 &
$m_{L_1}^{2}=m_{L_2}^{2}$ 
(TeV$ ^2$) & 
0.52
\\ \hline
$B_H~(\mbox{TeV}^2)$ &
-0.0027 &
$m_{e}^{2}=m_{\mu}^{2}$ 
(TeV$ ^2$) &
0.64
\\ \hline
$\mu_H$  (TeV) &
$\pm$ 0.94 &
$m_{Q_1}^{2}=m_{Q_2}^{2}$
(TeV$ ^2$) &
1.9 
\\ \hline
$m_{H_d}^{2}$ (TeV$ ^2$) &
-0.76 &
$m_{d}^{2}=m_{s}^{2}$
(TeV$ ^2$) & 
1.6
\\ \hline
$m_{H_u}^{2}$ (TeV$ ^2$) &
-0.90 &
$m_{u}^{2}=m_{c}^{2}$ 
(TeV$ ^2$) &
1.8 
\\ \hline
\end{tabular}

\vspace{0.2cm}

{\bf Table 1}:  Prediction of the SSB parameters.
\end{center}

\noindent
For the SSB parameters above we have used the notation of ref.
\cite{martin1}. Using these parameters, one can then compute
the superpartner spectrum.
We have checked that it is consistent 
with the experimental data.
The LSP, for instance, is found  to be 
a  neutralino of $\sim 220$ GeV
with a dominant component of the 
photino \footnote{The present example, however,  does not satisfy the
naturalness constraints \cite{dimopoulos1}.}.
Details of our calculations
and results will be presented elsewhere.

\vspace{0.3cm}

\noindent
We thank B. Ananthanarayan,
 M. Olechowski, R. Oehme,  K. Sibold, and
W. Zimmermann for useful discussions and suggestions.


\newpage
\begin{thebibliography}{99}

\bi{kmoz1} J. Kubo, M. Mondrag{\' o}n, M. Olechowski and
G. Zoupanos, \vspace{-0.2cm} {\em Unification of Gauge and
Yukawa Couplings without Symmetry}, hep-ph/9606434, 
\vspace{-0.2cm} to be
published in the Proc. of the {\em 5th Hellenic School
and Workshops on Elementary Particle Physics}, \vspace{-0.2cm}
Corfu, 3-24 September 1995.

\vspace{-0.3cm}
\bi{kmoz2} J. Kubo, M. Mondrag{\' o}n, M. Olechowski and
G. Zoupanos,  \vspace{-0.3cm} ``Testing Gauge-Yukawa
Unified Models by $M_t$'', hep-ph/9512435, 
 to be published in Nucl. Phys. B.

\vspace{-0.3cm}
\bi{nilles1} H.P. Nilles, Phys. Rep. {\bf 110} (1984) 1.

\vspace{-0.3cm}
\bi{jack1}I. Jack and D.R.T. Jones, Phys. Lett. {\bf B349}
(1995) 294; \vspace{-0.3cm}
I. Jack, D.R.T. Jones and K.L. Roberts, Nucl. Phys. {\bf B455}
(1995) 83.

\vspace{-0.3cm}
\bi{kazakov1}
D.I. Kazakov, M.Yu. Kalmykov, I.N. Kondrashuk
and  A.V. Gladyshev, Nucl. Phys. {\bf B471} (1996) 387.

\vspace{-0.3cm}
\bi{mondragon1} J. Kubo, M. Mondrag\'on and G. Zoupanos,
Nucl. Phys. {\bf B424} (1994) 291.

\vspace{-0.3cm}
\bibitem{zim1} W. Zimmermann,  Com. Math. Phys.
              {\bf 97} (1985) 211; \vspace{-0.3cm}
 R. Oehme and W. Zimmermann Com. Math. Phys.
              {\bf 97} (1985) 569.

\vspace{-0.3cm}
\bi{piguet1}
O. Piguet and K. Sibold, Phys. Lett. {\bf 229B} (1989) 83.

\vspace{-0.3cm}
\bi{zim2}
D. Maison, unpublished; W. Zimmermann, unpublished.

\vspace{-0.3cm}
\bi{sakai} S. Dimopoulos and H. Georgi, Nucl. Phys. {\bf B193}
(1981) 150; \vspace{-0.3cm}
 N. Sakai, Zeit. f. Phys. {\bf C11} (1981) 153.


\vspace{-0.3cm}
\bi{polonsky1} N. Polonsky and A. Pomarol, Phys. Rev. Lett.
{\bf 73} (1994) 2292.

\vspace{-0.3cm}
\bi{pendleton1}
B. Pendleton and G.G Ross, Phys. Lett. {\bf B98} (1981) 291.

\vspace{-0.3cm}
\bi{gates1} S.J. Gates, M. Grisaru,
M. Ro{\v c}ek and W. Siegel, \vspace{-0.3cm}
{\em Superspace or One Thousand and One
Lessons in Supersymmetry},
Benjamin/Cummins, 1983.


\vspace{-0.3cm}
\bi{lanross} M. Lanzagorta and G.G. Ross,  Phys. Lett. {\bf B349} (1995)
319; \vspace{-0.3cm}
P.M. Ferreira, I. Jack and D.R.T. Jones, Phys. Phys. {\bf B357}
(1995) 359.

\vspace{-0.3cm}
\bi{jack2}I. Jack, D.R.T. Jones and
G.G North, \vspace{-0.3cm} ``$N=1$ supersymmetry and
the three loop anomalous dimensions for the chiral superfield'',
hep-ph/9603386.

\vspace{-0.3cm}
\bi{ibanez1} L.E. Ib{\' a}{\~ n}ez and C. L{\' o}pez, Nucl. Phys. 
{\bf B233} (1984) 511.

\vspace{-0.3cm}
\bi{martin1} S.P. Martin and M. T. Vaughn, Phys. Rev. {\bf D50}
(1994) 2282.

\vspace{-0.3cm}
\bi{dimopoulos1}
S. Dimopoulos and G. Giudice, 
Phys. Lett. {\bf B357} (1995) 573.


\end{thebibliography}
\end{document}